# A Reliable SVD based Watermarking Scheme

Chirag Jain[1], Siddharth Arora[1], and Prasanta K. Panigrahi[1,2,†]

[1]*Physical Research Laboratory, Navrangpura, Ahmedabad - 380 009, India*
[2]*Indian Institute of Science Education and Research (IISER) Kolkata, Salt Lake, Kolkata - 700106, India*

**Abstract**

We propose a novel scheme for watermarking of digital images based on singular value decomposition (SVD), which makes use of the fact that the SVD subspace preserves significant amount of information of an image, as compared to its singular value matrix, Zhang and Li (2005). The principal components of the watermark are embedded in the original image, leaving the detector with a complimentary set of singular vectors for watermark extraction. The above step invariably ensures that watermark extraction from the embedded watermark image, using a modified matrix, is not possible, thereby removing a major drawback of an earlier proposed algorithm by Liu and Tan (2002).

## 1 Introduction

Recently, a SVD based watermarking scheme has been proposed by Liu and Tan (2002), which tries to take advantage of the optimal image decomposition property of SVD for embedding a watermark in an image. It was however argued by Zhang and Li (2005), that by taking recourse to the reference matrices of the watermark, the same can be extracted from a possibly distorted watermarked image. The fact that SVD subspace can preserve major information of an image, leads to the above-mentioned flaw, in which any reference watermark that is being searched for in an arbitrary image can be found.

The above mentioned problem has been dealt by Kundur and Hatzinakos (2004), where multi-resolution data fusion principle is employed for watermark hiding and retrieval, while ensuring that the probability of false positive detection remains low. Recently, a counterfeiting attack has also been proposed, along with its solution by Wu (2005), whereby an adversary can claim the ownership of an image by fabricating a bogus original image and a logo. The goal of this paper is to propose a minimal alteration of the scheme proposed by Liu and Tan [2] that avoids the above problems. Our scheme preserves all the positive aspects of SVD based watermarking schemes and ensures that the watermark image has high correlation with the extracted watermark. Desired properties of watermarking schemes have been discussed and stated previously, Cox et al. (1997), and Pitas (1996).

---

[1] †Email:prasanta@prl.res.in

Singular value decomposition is a general linear algebra technique, whereby a given matrix (image in this case), is diagonalized such that most of its signal energy is localized in a few singular values, Golub and Reinsch (1970). Hence, a digital image A of size M x N can be represented by its SVD as,

$A = USV^T,$

where U and V are orthogonal matrices of size M x M and N x N, respectively. S is a diagonal matrix of size M x N, with the diagonal elements representing the singular values (SV's). Columns of matrix U, also known as left singular vectors, are the eigenvectors of $AA^T$, while columns of matrix V (right singular vectors) are eigenvectors of $A^TA$. It is worth noting that, the singular vectors of an image specify the image 'geometry', while the singular values specify the 'luminance' (energy) of the image. It is found that slight variations in the singular values do not affect the visual perception of the quality of the image. It is this property, based on psycho-visual effect, that allows one to embed the watermark bits in the original image through minor modification of the singular values of the original image. However, it is worth noting that for an image of size M x N, the singular vectors have $O(N^2)$ elements, as compared to just N diagonal elements in the singular value matrix. Hence, this makes the use of singular vectors for information hiding more appropriate than using the singular values, as used by Chang et al. (2005) and also by Agrawal and Santhanam (2006). It is these elegant properties of SVD that has attracted attention of researchers for using it for watermarking [10-20].

Below, we briefly outline the algorithm proposed by Liu and Tan, and the flaws thereof, as pointed out by Zhang and Li. The present algorithm is proposed thereafter, with optimal modification of the one due to Liu and Zhang, using the inherent properties of SVD, demonstrating theoretically that our scheme is free of the flaws pointed above. We then provide the simulation results in Section II, followed by discussion and conclusion.

## 2  SVD based watermarking scheme

In the scheme proposed by Liu and Tan, the watermark is embedded in the original image in the singular value decomposition domain. Performing SVD on the original image 'A' yields,

$$A = USV^T, \tag{1}$$

A spread spectrum watermark image 'W' is then added to the above obtained singular values of A to get a new matrix $S + \alpha W$, where $\alpha$ is a scaling parameter. Obtaining the singular values of this new matrix through SVD, we get

$$S + aW = U_w S_w V_w^T \tag{2}$$

The watermarked image $A_w$ is then obtained using the eigenvectors of the original image as,

$$US_w V^T = A_w \tag{3}$$

During watermark extraction, the embedded watermark W* is obtained from a possibly distorted watermarked image $A_w^*$ by reversing the embedding process as follows,

$$A_w^* => U^* S_w^* V^{*T} \tag{4}$$

$$U_w S_w V_w^T => D^* \tag{5}$$

$$(D^* - S)/\alpha => W^* \tag{6}$$

However, it was later pointed out by Zhang and Li, that the addition of singular values S (diagonal matrix) to a scaled down version of watermark W (non-diagonal), results in a slightly modified W, whose sub-spaces differ from that of the original watermark only along the diagonal elements.

Since the matrices $U_w$, S and $V_w$ are assumed to be known for watermark detection, anyone who uses an arbitrary diagonal matrix along with known $U_w$, and $V_w$, will extract the watermark image with only the diagonal elements being different from the original watermark image. Secondly, the scheme is also vulnerable in the sense that any reference watermark that is being searched for, can be found out from an arbitrary image. This was beautifully demonstrated by Zhang and Li by embedding two watermarks in the original image A. The two watermarks 'Baboon' $W_b$, and watermark 'Plane' $W_p$ are embedded in the host image to generate two watermarked images $A_{wb}$ and $A_{wp}$ as specified in eq 1, 2, and 3. At the detector end, one does not know what the embedded image is. So searching for 'Plane' in the image that is watermarked with 'Baboon' yields the following result,

$$A^*_{wb} => U^*_{wb} S^*_{wb} V^{*T}_{wb} \tag{7}$$

$$U_{wp} S^*_{wb} V_{wp}^T => D^*_{mp} \tag{8}$$

$$(D^*_{mp} - S)/\alpha => W^*_{mp} \tag{9}$$

Ideally, the extracted watermark $W^*_{mp}$ should have no correlation with 'Plane' since the embedded image is that of 'Baboon'. But as the subspaces of original watermark and modified watermark differ only along the diagonal, majority of image information is preserved in SVD subspace leading to reference image retrieval ('Plane') from an arbitrary image ('Watermarked with Baboon'). Note that the diagonal elements of the reference image are obtained with error in the diagonal elements [Fig. 1f]; this is expected as the subspace of $S + \alpha W$ differs from W, only along the diagonal.

Taking into consideration the above flaws and the fact that the singular vectors U and V have majority of image information, we propose a scheme whereby the principal components of watermark image are embedded into the singular values of original image. Using this scheme, only a set of singular vectors are required to be known at the detector. The main steps of our proposed scheme are as follows,

$$A = USV^T, \tag{10}$$

$$W => U_w S_w V_w^T \tag{11}$$

$$=> A_{wa} V_w^T$$

where A is the original image and W is the watermark to be embedded in A. $A_{wa} = U_w S_w$ are also known as principal components. Embedding the principal components $A_{wa}$ with a diagonal singular value matrix S of the original image, we get $S_1$. From $S_1$ we get the watermarked image $A_w$ as follows,

$$S_1 = S + \alpha A_{wa} \tag{12}$$

$$US_1 V^T => A_w \tag{13}$$

Let $A^*_w$ denote the possibly distorted watermarked image at the detector. To recover back the watermark image $W^*$ from $A^*_w$, following are the steps involved,

$$(A^*_w - A) => A_1 \tag{14}$$

$$(U^{-1} A_1 (V^T)^{-1})/\alpha => A^*_{wa} \tag{15}$$

$$A^*_{wa} V^T_w => W^* \tag{16}$$

We now try to search for a reference image, e.g., 'Plane', in the image watermarked with baboon using our scheme. We first find the SVD of the original plane image P as,

$$P => U_p S_p V_p^T$$

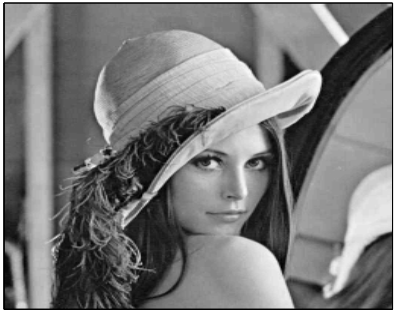
(a)

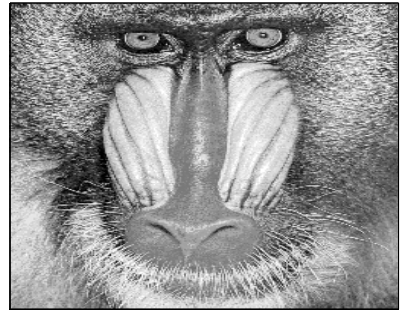
(b)

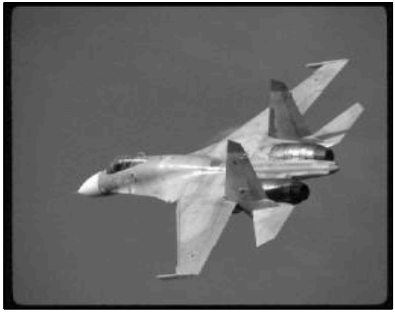
(c)

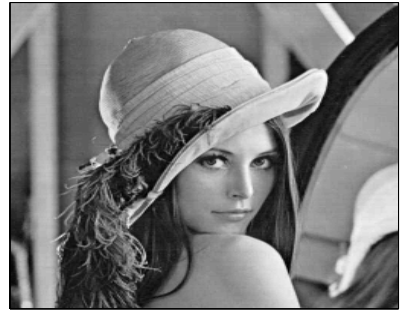
(d)

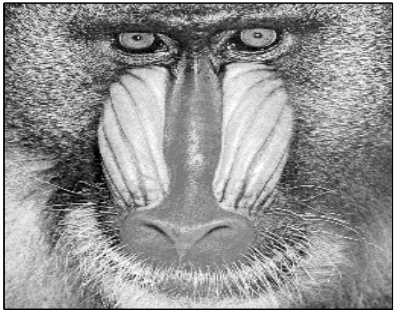
(e)

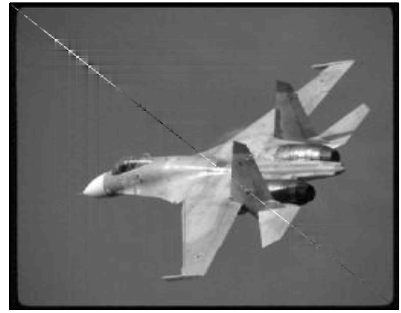
(f)

Fig. 1. (a) Original Lena image (b) Original Baboon image (c) Original Plane image (d) Watermarked image, baboon embedded in Lena image (e) Watermark extracted from 1d using our scheme (f) Plane Extracted from watermarked image using Liu and Tan's scheme

To find a possibly distorted plane image in the distorted watermarked image $A^*_{wa}$, we use the singular vectors of reference image as follows,

$$P^* => A^*_{wa} V_p^T \qquad (17)$$

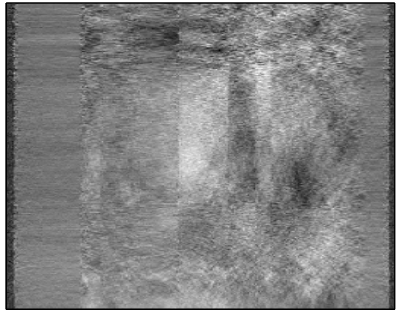

Fig. 2. Distorted reference image 'Plane' extracted from fig. 1d using our scheme.

Since, one of the singular vectors of watermark is embedded in the original image; watermark extraction without knowing the original principal components is not possible. Hence, no reference image can be extracted from any arbitrary image using our scheme [Fig.2].

## 3  Conclusion

In this paper we have presented a singular value based watermarking scheme, where we embed the principal components of the watermark in the original image rather than just the singular values. The fact that the principal components have been added to the singular values of original image achieves two useful purposes. Firstly, the information about the entire watermark is not available without a prior knowledge of the original watermark. This is of significance for the security of the watermark. Secondly, the method avoids the pitfall encountered by Liu and Tan, where the watermark was modified only along the diagonals, leading to the extraction of a reference watermark that is being searched using an arbitrary image. Hence, our method utilizes the property of SVD based watermarking algorithms and ensures rightful ownership of the digital watermark image.